\documentclass[aps,prb,twocolumn,superscriptaddress]{revtex4-1}

\usepackage{physics}
\usepackage{graphicx}
\usepackage{dsfont}
\usepackage{xcolor}
\usepackage[utf8]{inputenc}

\newcommand{\up}{\uparrow}
\newcommand{\dn}{\downarrow}

\newcommand{\refeq}[1]{Eq.~(\ref{#1})}
\newcommand{\refsec}[1]{Section~(\ref{#1})}

\newcommand{\reffig}[1]{Fig.~\ref{#1}}

\newcommand{\eV}{\mathrm{eV}}

\def\hatc{\hat{c}}
\def\cc{\hat{c}^{\dagger}}
\def\n{\hat{n}}
\def\s{\sigma}
\def\a{\alpha}

\begin{document}

\title{Electronic correlations and competing orders in multiorbital dimers:\\ a cluster DMFT study}
\author{Malte~Harland}
\affiliation{Institute of Theoretical Physics, University of Hamburg, Jungiusstra{\ss}e 9, 20355 Hamburg, Germany}
\author{Alexander~I.~Poteryaev}
\affiliation{Institute of Metal Physics, S. Kovalevskoy 18, 620219 Ekaterinburg GSP-170, Russia}
\author{Sergey~V.~Streltsov}
\affiliation{Institute of Metal Physics, S. Kovalevskoy 18, 620219 Ekaterinburg GSP-170, Russia}
\affiliation{Ural Federal University, Mira St. 19, 620002 Ekaterinburg, Russia}
\author{Alexander~I.~Lichtenstein}
\affiliation{Institute of Theoretical Physics, University of Hamburg, Jungiusstra{\ss}e 9, 20355 Hamburg, Germany}
\affiliation{Ural Federal University, Mira St. 19, 620002 Ekaterinburg, Russia}
\date{\today}

\begin{abstract}
  We investigate the violation of the first Hund's rule in 4$d$ and 5$d$ transition metal oxides that form solids of dimers. Bonding states within these dimers reduce the magnetization of such materials. We parametrize the dimer formation with realistic hopping parameters and find not only regimes, where the system behaves as a Fermi liquid or as a Peierls insulator, but also strongly correlated regions due to Hund's coupling and its competition with the dimer formation. The electronic structure is investigated using the cluster dynamical mean-field theory for a dimer in the two-plane Bethe lattice with two orbitals per site and $3/8$-filling, that is three electrons per dimer. It reveals dimer-antiferromagnetic order of a high-spin (double exchange) state and a low-spin (molecular orbital) state. At the crossover region we observe the suppression of long-range magnetic order, fluctuation enhancement and renormalization of electron masses. At certain interaction strengths the system becomes an incoherent antiferromagnetic metal with well defined local moments.
\end{abstract}
\pacs{}
\maketitle

\section{Introduction}
A standard paradigm of the strongly correlated materials involves competition between on-site Coulomb repulsion ($U$), which tends to localize electrons on particular sites, and band effects (characterized, e.g., by the width of a corresponding band, $W$) making them delocalized\cite{Mott1949,Imada1998}. In effect there can be a transition from a homogeneous metallic to a homogeneous insulating state. In real materials this picture can be enriched by an unusual band topology\cite{Pesin2010,Witczak-Krempa2014}, spin-orbit coupling\cite{Kim2008,Witczak-Krempa2014}, interplay between different degrees of freedom such as orbital, charge, spin etc.\cite{KK-UFN,Tokura2000,khomskii2014transition} However, there can be another option - a system may prefer an inhomogeneous scenario forming metallic clusters within an insulating media (molecules-in-solids conception\cite{Streltsov2017}). The simplest example of such clusters is a dimer. If $U$ is not very large the wavefunction is essentially a molecular orbital with an electron delocalized over both sites. But there are also materials with other types of clusters: trimers\cite{Takayama2014,Sheckelton2017}, tetramers\cite{Radaelli2002} and even heptamers\cite{Horibe2006}. The electrons can easily propagate within these clusters, but hoppings between them are suppressed.

There are two main problems in this concern. First of all, there is no general theory, which explains why some of the systems stay homogeneous, while others form (spontaneously) clusters. We knew for a long time that such transitions can be induced by Peierls and spin-Peierls effects\cite{Bulaevskii1975,Kagoshima1981}, or, more generally, by a charge density wave (CDW) instability due to nesting of the Fermi surface\cite{khomskii2014transition,Chen2016}, but a complete understanding of how strong electronic correlations, spin-orbit and exchange couplings affect this transition is still lacking. Moreover, calculations for real materials show that there is no nesting in many systems, whose properties were supposed to be explained by the formation of a CDW, or that there is nesting at a wrong wave vector\cite{Johannes2008a}.

Another problem is a theoretical description of such inhomogeneous systems. While the homogeneous situation with Mott-Hubbard transition was extensively investigated over the years, physical properties of clusterized materials still remain mostly unexplored. Up to now most of the efforts were concentrated on the study of the so-called two plane Hubbard model (known also as dimer Hubbard model), which is the Hubbard model on the Bethe lattice composed of dimers, see also Sec.~\ref{sec:methods}.  Most of the attention has been paid to the situation with one orbital per each site in a dimer and half-filling\cite{Moeller1999,Fuhrmann2006,Kancharla2007,Hafermann2009}. This model allows to describe the transition from a band to Mott insulator and is particularly relevant for such materials as VO$_2$, V$_2$O$_3$, and Ti$_2$O$_3$\cite{Poteryaev2004,Biermann2005,Najera2017,Najera2018}. The two-orbital case has been considered for the one-dimensional chain using the dynamical mean-field theory (DMFT).\cite{Streltsov2014,Streltsov2016} The orbital-selective behaviour has been found for different electron fillings and has been shown to strongly affect magnetic properties of a system, since some of the electrons occupying bonding orbitals may form spin-singlets. In effect, only a part of the electrons contribute to the total magnetic moment. This violates the Hund's rules and may dramatically change exchange coupling between neighbouring dimers\cite{Streltsov2016}. However, the one-dimensional lattice is not a natural choice for the DMFT because of the small number of nearest neighbors.

Hund's coupling stems from the Coulomb interaction, it represents the intra-atomic exchange and has strong influences on the electronic correlations and therefore also on the Mott transition.\cite{Georges2013,Medici2011a} It can shift the critical interaction value of the Mott transition and also diminish or promote the coherence of Fermi-liquids. This depends strongly on the filling\cite{Medici2011}, i.e. for half-filling the effective Coulomb interaction is increased and for all other fillings it is decreased. Therefore, Hund's coupling can suppress the Mott transition, but not the correlations. Thus there can be strongly correlated materials, that are not close to a Mott transition, but still exhibit enhanced electron masses, local moments and orbital-selectivity.\cite{Pavarini2017}

In the present paper the simplest model of multiorbital (two orbitals) dimers on the two plane Bethe lattice with the odd number of electrons (three) is considered. The parameters of the model are chosen to be close to those specific parameters in real materials based on the late transition metal ions. We not only find the transition between states with different total spin ($S=1/2$ to $S=3/2$) as a function of the hopping in the dimer, but also observe the suppression of the long-range magnetic ordering by the temperature in the crossover region nearby this transition. Moreover, surprisingly such a transition can be induced by the hopping in the Bethe lattice. We discuss the electronic and magnetic properties of the considered two-plane Bethe lattice model and identify regimes where the system behaves as a Fermi liquid, a Peierls insulator and a correlated metal. These results not only advance our knowledge on the properties of the two plane Bethe lattice model, but also can be useful for the description of dimerized materials, which are under close investigation nowadays.

\section{Model  \& Method\label{sec:methods}}
\begin{figure}
  \centering
  \includegraphics[scale = .85]{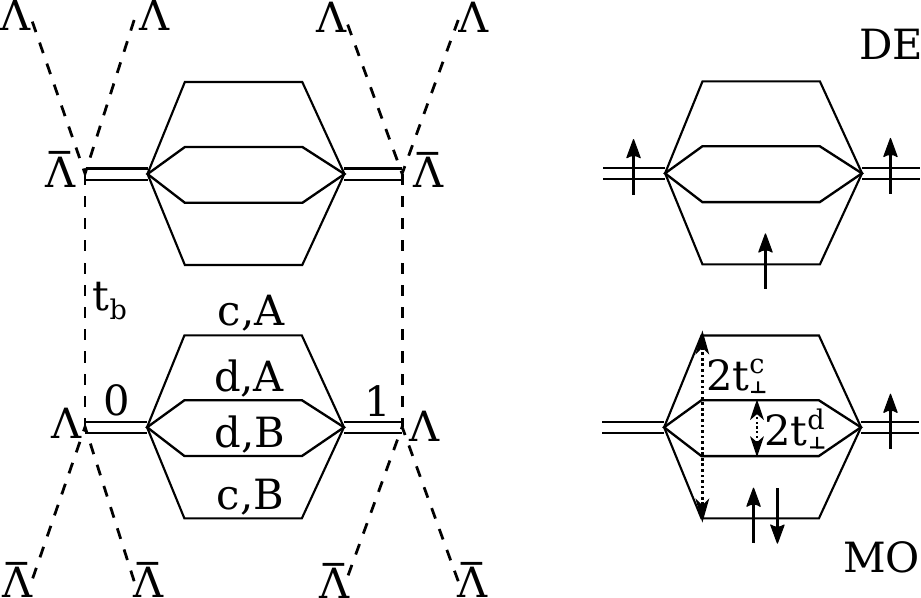}
  \caption{Left: The two Bethe lattices with hopping $t_b$ are interconnected by dimers of two atomic orbitals ($c$, $d$) and two sites ($0$, $1$). The Bethe lattice coordination is finite, i.e. $z=3$, for illustrative purposes. The sites can form bonding($B$) and antibonding($A$) molecular orbitals. The bipartite Bethe lattice can be divided into sublattices ($\Lambda$, $\bar{\Lambda}$). Right: Two possible ground state configurations in case of $N=3$ electrons: the molecular-orbital (MO) and the double exchange (DE) states. Their  competition is defined by the Hund's exchange coupling $J$, screened intra-($U$) and interorbital ($U^\prime$) Coulomb repulsion and the dimer hoppings $t^c_\perp$ and $t^d_\perp$.}
  \label{fig:model}
\end{figure}
While the two-plane Hubbard model seems to be a rather natural choice in the case of VO$_2$ with a single electron in the $3d$ shell, for a realistic description of materials with larger number of $d$ electrons one needs to take into account the orbital degeneracy and possible crystal field splitting. The latter can be due to {\it i)} a nearest neighbour ligand's environment (below for simplicity we will consider the octahedral case) and {\it ii)} because of bonding with other transition metal ions.

 The dimerization occurs, when two transition metal ions are able to come close enough to each other to lower the total energy due to formation of bonding orbitals. This is possible, when ligand octahedra share their edges or faces, whereas a common corner geometry prevents dimerization because of negatively charged ligand sitting in between of two transition metals. Edge-sharing structures can be met, e.g., in delafossites, spinels, very popular now 213 honeycomb iridates and ruthenates. Face-sharing is more common in one-dimensional materials like 6H-perovskites, ZrI$_3$ etc., but there are also three-dimensional corundum-like structures.

It is rather important that in addition to trivial splitting of the $d$ orbitals onto lower-lying $t_{2g}$ and higher-lying $e_g$ manifolds there is always an additional splitting in these geometries due to neighboring transition metal ions. The last can be effectively integrated out. In the edge-sharing octahedra the $t_{2g}$ orbitals turn out to be split onto the $xy$ and $yz/zx$ orbitals. The $xy$ orbitals of neighboring metals point to each other. This results in a strong bonding-antibonding splitting, while the $xz/yz$ orbitals may be still considered as site-localized\cite{Streltsov2017}. This is especially important for the $4d$ and $5d$ transition metal ions, since their wavefunctions are more spatially extended and corresponding bonding-antibonding splitting is much larger than for the $3d$ transition metal ions. A similar situation occurs for face-sharing octahedra, where the $a_{1g}$ orbitals form a bonding orbital and the $e^\pi_g$ orbitals remain localized\cite{Kugel2015,Khomskii2016}.

Thus, in order to describe dimerized transition metal compounds with more than one electron one needs at least two different sets of atomic orbitals, which differ by value of the hopping parameters. Due to computational limitations we will restrict ourselves to the minimal model with two orbitals per site. We label the orbital forming the molecular-orbital $c$ and the localized one $d$ (see \reffig{fig:model}). Corresponding intradimer hopping parameters are $t^c_\perp$ and $t^d_\perp$. A dimer is considered as a vertex of the Bethe lattice with infinite coordination. For simplicity we assume that hoppings along the Bethe lattice, $t_b$, is the same for both orbitals.  Spatial correlations beyond the dimer don't exist because of the infinite coordination.

The Hamiltonian of the model above is
\begin{equation}
  \hat{H} =  -t_b \sum_{\langle \lambda,\lambda' \rangle \s} \sum_{\a i}   \cc_{\lambda \s \a i} \hatc_{\lambda' \s \a i}
                + \sum_{\lambda} \hat{H}_{\lambda}^{Dimer},
  \label{eq:Hamiltonian}
\end{equation}
where $\l$ denotes a nearest-neighbour dimers, $\s$ is a spin, $i=\{0,1\}$ runs over sites within a dimer and $\a=\{c,d\}$ is an orbital index. Therefore, the first term describes a hopping of the electron between dimers with the amplitude $t_b$ and the second term is responsible for the ``local'' (intra-dimer) interaction and can be written as
\begin{align}
  \hat{H}_{\lambda}^{Dimer} & =  \sum_{\s i\a} t_{\perp}^{\a}   \cc_{\lambda\s\a i} \hatc_{\lambda\s\a\bar{i}}
                         + U   \sum_{i\a} \n_{\lambda\up\a i} \n_{\lambda\dn\a i} \nonumber\\
                       & + U' \sum_{\s i} \n_{\lambda\s c i} \n_{\lambda\bar{\s} d i}
                         + ( U' - J )  \sum_{\s i} \n_{\lambda\s c i} \n_{\lambda\s d i} \nonumber\\
  - J \sum_{i} & (   \cc_{\lambda\dn c i} \cc_{\lambda\up d i} \hatc_{\lambda\dn d i} \hatc_{\lambda\up c i}
                 + \cc_{\lambda\up d i} \cc_{\lambda\dn d i} \hatc_{\lambda\up c i} \hatc_{\lambda\dn c i} + \mathrm{h.c.}  ).
  \label{eq:H_dimer}
\end{align}
The orbital differentiation (the first term) is caused by the intradimer hopping parameters, $t_{\perp}^{\a}$ and we do not introduce crystal-field splittings ($c$-$d$). The intradimer hopping can also be written in matrix notation
\begin{equation}
  \label{eq:tloc}
  t_{loc} = \begin{pmatrix} -t^c_\perp & 0\\ 0 & -t^d_\perp \end{pmatrix} \otimes \sigma_x,
\end{equation}
where the Pauli matrix $\sigma_x$ creates the off-diagonal entries of the site-space. The local electron-electron interaction at each site (the last terms in \refeq{eq:H_dimer}) is modelled via the Kanamori parametrization\cite{Kanamori1963}, where  $U, U^\prime$ are intra-/inter- orbital Coulomb repulsions and $J$ is the Hund's exchange coupling. We choose the interorbital Coulomb interaction by cubic symmetry as $U^\prime = U - 2J$.

The model is solved at finite temperatures exactly using the cluster dynamical mean-field theory (CDMFT)\cite{Georges1996,Kotliar2004,Maier2005,Kotliar2006} with a continuous-time quantum Monte-Carlo impurity solver (CTHYB)\cite{Gull2011,Boehnke2011,Seth2016}. The solver as well as the CDMFT code have been written using the TRIQS library\cite{Parcollet2015}.

The dimer's degrees of freedom of our auxiliary impurity model contain two spins, two orbitals and two sites. The Bethe lattice can be divided onto two equivalent sublattices $\Lambda$ and $\bar{\Lambda}$, see \reffig{fig:model}. The CDMFT self-consistency equation describes a particle of $\Lambda$ fluctuating through its environment $\bar{\Lambda}$. Since we are interested in a solution of a broken spin-symmetry, we apply the antiferromagnetic condition for the construction of the Weiss field,
\begin{equation}
  \label{eq:dmft}
  \mathcal{G}^{-1}_\sigma(i\omega_n) = (i\omega_n + \mu)\mathds{1} - t_{loc} - t^2_b G_{-\sigma}(i\omega_n),
\end{equation}
where $\mathds{1}$ is unit matrix, $\mathcal{G}(i\omega_n)$ is the Weiss field and $G(i\omega_n)$ is the local Green's function,  both are matrices in spin, orbital, and site space. Note, that the antiferromagnetic order described by \refeq{eq:dmft} exists between the dimers (dimer-antiferromagnetism) and not within them. To find CDMFT solutions of broken spin symmetry we add a small external magnetic field to the Hamiltonian, which is switched off after a few CDMFT iterations. It is worth mentioning that there are also other interesting solutions, which allow for the coexistence of insulating behaviour and ferromagnetism\cite{Nishimoto2012}, but the study of this part of the phase diagram lies beyond scope of the present paper. We also use a diagonal-basis of the site-space in the blockstructure of the Green's function (see below), and thereby solutions of broken site symmetry within dimers are excluded, i.e. the charge ordering within the dimers was forbidden by construction.

The local Green's function, which is needed to calculate the chemical potential $\mu$ in course of the CDMFT self-consistency, can be found using following equation:
\begin{equation}
  \label{eq:gloc}
  G^{-1}_\sigma(i\omega_n) = (i\omega_n + \mu)\mathds{1} - t_{loc} - t^2_b G_{-\sigma}(i\omega_n) - \Sigma_\sigma(i\omega_n).
\end{equation}
This implicit equation has to be solved iteratively starting with setting it equal to the impurity Green's function of the last CDMFT cycle $G(i\omega_n) = g(i\omega_n)$, which is also the self-consistency condition for the CDMFT. The self-energy is calculated via Dyson equation from the impurity quantities $\Sigma(i\omega_n) = \mathcal{G}^{-1}(i\omega_n) - g^{-1}(i\omega_n)$ and initially is set to zero. 

In order to make the quantum Monte-Carlo impurity solver more efficient we use a standard unitary transformation on the site-space $j \in \{0, 1\}$:
\begin{equation}
  \label{eq:unitarytransf}
  {\hat {\tilde c}}_{\s \a i} = \Sigma_{i} T_{ij} \hatc_{\s \a j},\quad   T = \frac{1}{\sqrt{2}}
  \begin{pmatrix}
    1 && 1  \\
    1 && -1 \\
  \end{pmatrix}.
\end{equation}
transforming to the bonding ($B$)/antibonding ($A$) basis, labeled by $i \in \{A, B\}$, with corresponding creation/annihilation operators labeled by tilde. This transformation diagonalizes the local Green's function in site-space and thereby also in all single-particle orbitals.

To sum up, even taking into account all constraints and simplification there are still five parameters in our model ($U$, $J$, $t^c_\perp$, $t^d_\perp$, $t_b$). To reduce this number further we will restrict ourselves by typical values met in real materials. We choose two groups of compounds with the general formulas Ba$_3$MeTM$_2$O$_9$\cite{Kimber2012,Senn2013a,Panda2015,Feng2017,Ziat2017} and Re$_5$TM$_2$O$_{12}$\cite{Torardi1985,Jeitschko1999,Chi2003}, where Re is a rare-earth, TM is a transition metal ion, Me is a rare-earth, alkali or transition metal ion. There are dimers formed by two TMO$_6$ octahedra in these two classes of systems (sharing their faces in Ba$_3$MeTM$_2$O$_9$ and edges in Re$_5$TM$_2$O$_{12}$).

Typically TM ions are $4d$ metals such as Ru, Re, Mo, and Os for which Hund's exchange $J \sim 0.7\eV$ and Hubbard $U \sim 4.5\eV$ (i.e. $U'  \sim 3\eV$)\cite{Streltsov2017}. Therefore, we will fix the screened Coulomb interaction and Hund's exchange to the values above. The hopping of the more localized orbital is set to $t^d_\perp = 0.2\eV$. The hopping parameters, $t^c_\perp$ and $t_b$, will be varied in what follows. The density functional theory suggests also typical values of the hopping parameters in these materials: $t^c_\perp$ changes  from $0.7$ to $1.4\eV$, while $t^d_{\perp} \sim 0.2\eV$ and $t_b \sim 0.2\eV$\footnote{S.V. Streltsov to be published}. One should also note that the electron filling per dimer will be fixed to the value of $N=3$, i.e. $3/2$-electrons per site. Additionally, we would like to remind that the AFM and PM self-consistency conditions will be used over this study.

\subsection{Atomic limit at $T=0$\label{sec:atomic-limit}}
We start an analytic treatment of our model in the atomic limit, where the hopping in the Bethe planes is suppressed, i.e. $t_b = 0$. There are two possible ground states for the isolated dimer with $N=3$ electrons and two different orbitals $c$ and $d$. The first state with one electron residing bonding $c$ and other two electrons occupying $d$ orbitals with the same spin will be referred as double exchange (DE) configuration, since it maximizes total spin of the dimer. Another configuration, called molecular-orbital (MO) state, is the state with completely filled bonding $c$ orbital and remaining electron distributed over localized $d$ orbitals (the charge ordering is forbidden by construction, see Sec.~\ref{sec:methods}).

Neglecting quantum fluctuations one may approximate these states by the following wave functions:
\begin{align}
  \begin{split}
    \ket{MO} &= \cc_{\up d 0} {\hat {\tilde c}}_{\up c B}^{\dagger} {\hat {\tilde c}}^{\dagger}_{\dn c B} \ket{0},\\
    \ket{DE} &= \cc_{\up d 0} \cc_{\up d 1} {\hat {\tilde c}}^{\dagger}_{\up c B} \ket{0}.
  \end{split}
\end{align}
Here, we use the bonding-/antibonding basis only for the $c$ electrons, whereas the $d$ electrons are described by the site-localized basis and $\ket{0}$ is the vacuum state.
\begin{figure}
  \centering
  \includegraphics[scale=0.35]{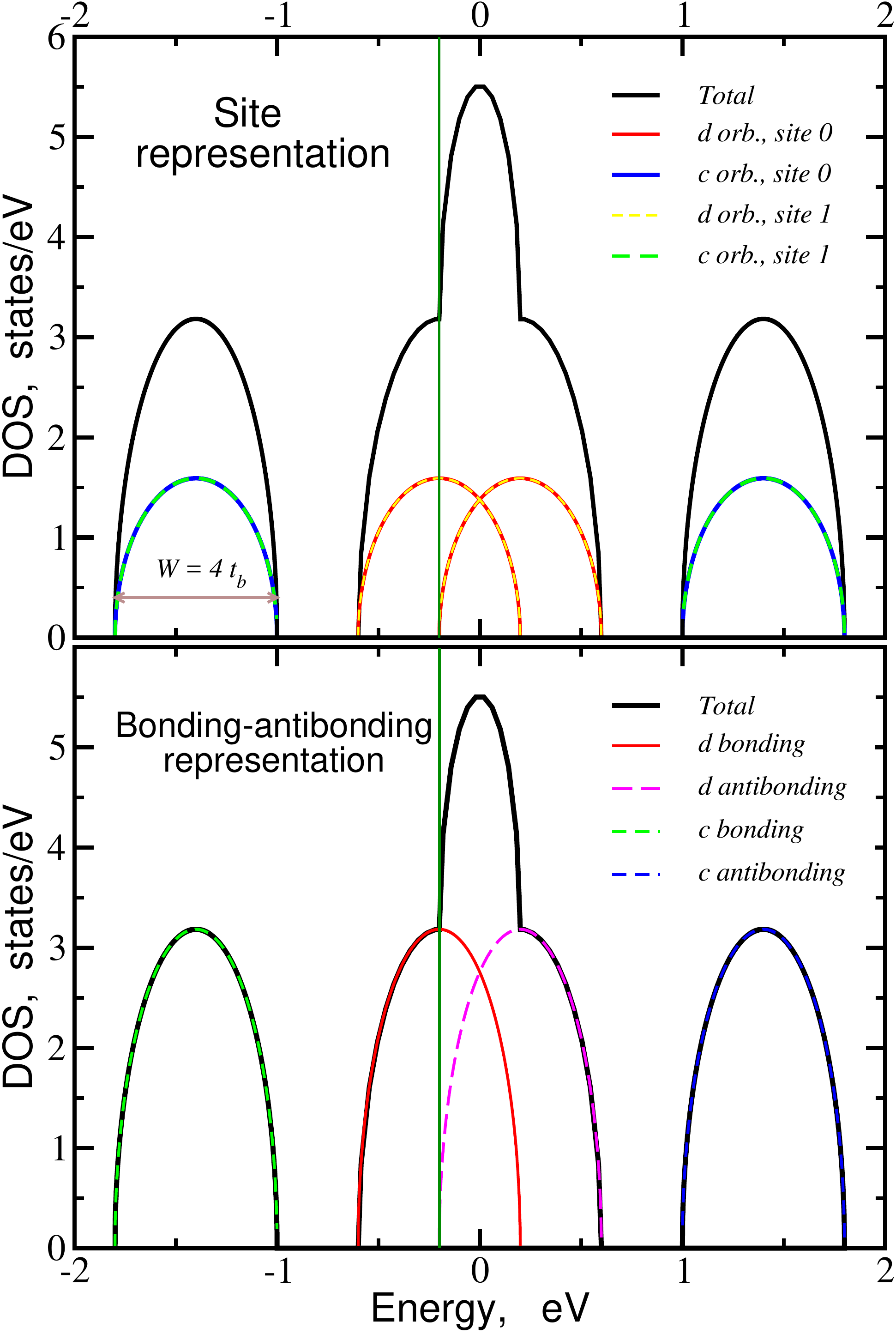}
  \caption{(Color online) The density of states in the non-interacting limit (i.e. $U=J=0$), as a function of intra-dimer hopping of the $c$-electrons $t^c_\perp$ and excitation energy $\omega$. Hopping within Bethe plane is chosen to be $t_b=t^d_\perp=0.2\eV$.}
  \label{fig:dossemicirculars}
\end{figure}
The energies of the DE and MO states can be easily estimated if one considers the density-density interactions of original Hamiltonian \eqref{eq:H_dimer}. Since $t^c_{\perp} \gg t^d_{\perp}$ we neglect it in the first approximation. Thus, the energies of MO and DE states are:
\begin{align}
  \begin{split}
    E^{MO} &= \frac{11}{4} U - \frac{7}{4} J -  2t^c_\perp,\\
    E^{DE} &= 2 U - 5 J - t^c_{\perp}.
  \end{split}
\end{align}
Then the critical $\tilde{t}^c_\perp$ for the transition from the MO to DE configuration is
\begin{align}
  \label{eq:crittc}
  \tilde{t}^c_\perp = \frac{3U - 13J}{4},
\end{align}
which gives $\tilde{t}^c_\perp = 1.1\eV$ for our set of parameters ($U=4.5\eV$, $J=0.7\eV$).

\subsection{Non-interacting regime}
\reffig{fig:dossemicirculars} presents the electronic structure in the non-interacting regime. It is rather trivial and reminds the simplified sketch shown in the right part of Fig.~\ref{fig:model}. The density of states in this limit is a superposition of four semicirculars with the individual bandwidth $W=4t_b$. 
The bands, corresponding to the $c$ ($d$) orbitals, are centered at the energies of $\pm t^c_\perp$ ($\pm t^d_\perp$) in a site representation (see upper panel of
\reffig{fig:dossemicirculars}). A site equivalence leads to an overlay of the DOSes from different sites.
Since, in our consideration $t^c_\perp > t^d_\perp$, the $c$ bands are always further away from the Fermi level then the $d$ bands. One should note that the Fermi level is not at the middle of the $d$ band since we are not at the half-filling (which would be for $N=4$). 

The transformation of the non-interacting model to the bonding-antibonding (BA) representation simplifies drastically an examination of the DOS (see lower panel of \reffig{fig:dossemicirculars}). For example, in the site representation the band of $c$ character at site $0$ or $1$ was located at $-t^c_\perp$ and  $+t^c_\perp$. While, after BA transformation, there are two bands (instead of two sites) of pure $c$ bonding character at $-t^c_\perp$ and $c$ antibonding character at $+t^c_\perp$. Thus, in ascending order, one have four bands of pure character: $c$ bonding ($c,B$), $d$ bonding ($d,B$), $d$ antibonding ($d,A$) and $c$ antibonding ($c,A$).
The bonding and antibonding states are separated by the $2t^c_\perp-4t_b$ ($2t^d_\perp-4t_b$). If $t^c_\perp > t^d_\perp+4t_b$, there is a gap between $c$ (anti)bonding and $d$ (anti)bonding states. Additionally, if $t^d_\perp>2t_b$, there is a small gap between bonding and antibonding states of $d$ character. Formation of these bands can be considered as a local crystal field effect with the $t^c_\perp$ ($t^d_\perp$) playing a role of a crystall field splitting. 

Crystal fields are known to compete with the Hund's coupling. This leads to a number of very important phenomena, such as, e.g., spin-state transitions\cite{khomskii2014transition}. Whereas intra-atomic Hund's exchange tends to the uniform orbital occupancy (strictly speaking this can be achieved only at half-filling), the crystal field promotes orbital polarization, when some of the orbitals are less occupied than others\cite{Lechermann2005}. But contrary to the conventional crystal field the formation of molecular orbitals also substantially changes the interaction terms. In the next section we discuss phase diagram of the two-plane Bethe lattice for an intermediate situation, when
both intradimer hoppings and interaction (given by $U$ and $J$) strength are not small.

\section{Phase diagram}
\begin{figure}
  \centering
  \includegraphics{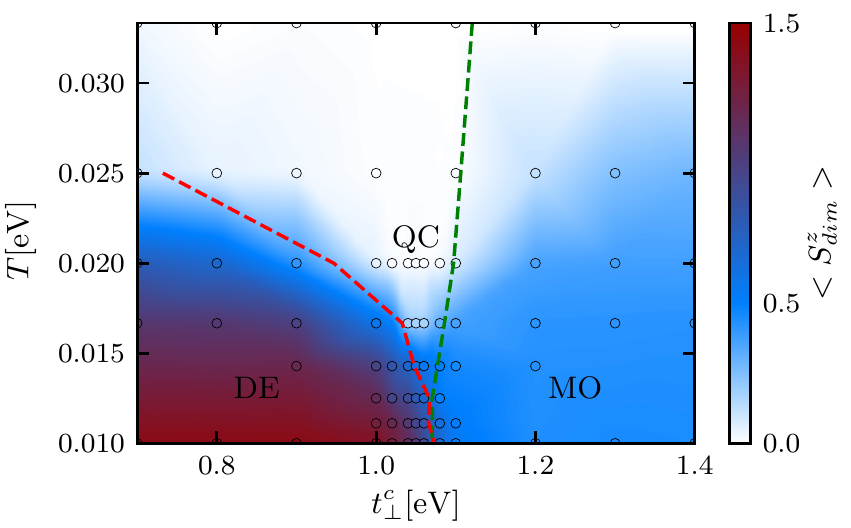}
  \caption{(Color online) The magnetic moment of the dimer $\expval{S^z_{dim}}$ as a function of temperature $T$ and intra-dimer hopping of the $c$-orbitals $t^c_\perp$ for $t^d_\perp=t_b = 0.2\eV$. The red and green dashed lines mark the positions of local minima of the spin and orbital correlations, respectively.}
  \label{fig:sztemptctd02}
\end{figure}
Previous studies of the two-plane Bethe lattice have focused on the single orbital case. It was found to hold not only the Mott and band insulators, but also a correlated mixed state with coherent and incoherent peaks in the local density of states. Competition between intra and interplane exchange interactions was shown to affect the formation of the local moments.\cite{Moeller1999,Hafermann2009,Najera2018} We will demonstrate that substantial orbital differentiation due to different interplane hoppings, $t^c_\perp \gg t^d_\perp$, results not only in a spin-state-like transition, but also in a strong suppression of a long-range magnetic order in the critical region.

Throughout this section we discuss results for fixed $t_b = 0.2\eV$. \reffig{fig:sztemptctd02} shows the phase diagram of our model. There are three main regions.  At low temperature and for small  $t^c_\perp$ we find the DE state with a total spin $S^z_{dim} = 3/2$ (red part of the phase diagram). All dimers are antiferromagnetically ordered, so that $\expval{S^z_{dim}} \sim 3/2$.  This DE state transforms into the MO state with the total spin $S^z_{dim}=1/2$ upon increasing intra-dimer hopping $t^c_\perp$ (light blue part of the phase diagram). This can be considered as a spin-state transition for the cluster. The critical $  \tilde{t}^c_\perp$ is close to the value obtained in the atomic limit (see Sec.~\ref{sec:atomic-limit}). At low temperatures dimers in the MO phase are antiferromagnetically ordered and $\expval{S^z_{dim}} \sim 1/2$.

Increasing the temperature we get to the last region with paramagnetic dimers (this phase can be again divided according to $\expval{S^2_{dim}}$ in the DE or MO parts). Interesting is, however, the fact that the temperature dependence of $\expval{S^z_{dim}}$ is very different in different parts of the phase diagram. We see that the paramagnetic phase appears at much lower temperatures in the critical region of $t^c_\perp \sim 1.05\eV$. The DE and MO states have different quantum numbers (different total spins) and thus in the limit of isolated dimers ($t_b=0$) the transition between them must be discontinuous at $T=0$. Obviously, no long-range magnetic order is possible in this situation. However, fluctuations can result in a crossover. In this crossover region the system becomes frustrated and the paramagnetic phase is promoted by the competition of the DE and MO states forming a hybrid state (HYB) with properties that are distinct from both.

\begin{figure}
  \centering
  \includegraphics{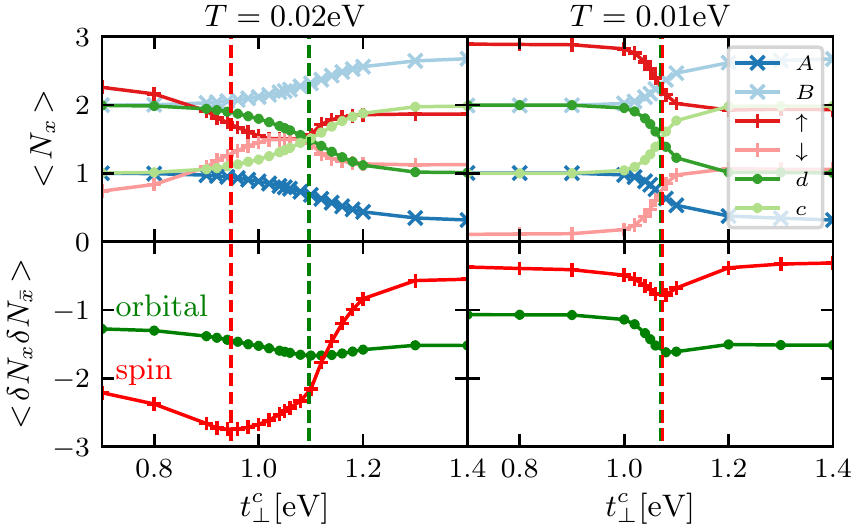}
  \caption{(Color online) Occupations (top) and correlations (bottom) of the single-particle orbitals across the $t^c_\perp$ driven DE/MO transition at $T=0.02\eV$ (left) and $T=0.02\eV$ (right). $N_x$ is the integrated occupancy. Dashed lines mark the $t^c_\perp$ of the correlation's respective local minimum for the spin (red)- and orbital (green)-correlations.}
  \label{fig:staticobs}
\end{figure}
The integrated occupancies 
\begin{align}
  \label{eq:Ni}
  \begin{split}
    N_\sigma &= \sum_{\alpha \in \{ c, d \}} \sum_{i \in \{ B, A\}} {\tilde n}_{\sigma \alpha i},\quad \sigma \in \{ \up, \dn \},\\
    N_\alpha &= \sum_{\sigma \in \{ \up, \dn \}} \sum_{i \in \{ B, A\}} {\tilde n}_{\sigma \alpha i},\quad \alpha \in \{ c, d\},\\
    N_i &= \sum_{\sigma \in \{ \up, \dn \}} \sum_{\alpha \in \{ c, d\}} n_{\sigma \alpha i},\quad i \in \{ B, A\},
  \end{split}
\end{align}
are shown in \reffig{fig:staticobs} (top), it confirms our illustration of the DE and MO states (\reffig{fig:model}). For low temperatures fluctuations are suppressed by the AFM order and the integrated occupancy has a sharper crossover. In fact the crossover region shows in proximity of its boundaries local minima of the spin and orbital correlations $\expval{\delta N_x \delta N_{\bar{x}}} = \expval{N_x N_{\bar{x}}} - \expval{N_x} \expval{N_{\bar{x}}}$ with $x = \up, \dn$ and $x = d, c$, respectively. The physical reasoning behind this is that the fluctuations are always very strong in vicinity of phase transitions. The temperature dependence of the $\expval{\delta N_x \delta N_{\bar{x}}}$ minima are shown in \reffig{fig:sztemptctd02} by dashed lines.

The phase diagram shows, that both originate from the DE/MO groundstate crossover, but their temperature dependence is very different. The spin correlation minimum is very close to the critical temperature of the DE state for all $t^c_\perp$. The decoupling of spins is much stronger than that of the orbitals, which is rather independent of the temperature. The comparison of the correlations at different temperatures (\reffig{fig:staticobs} left and right) shows, that also the magnitude of spin fluctuations of the DE state depends strongly on the temperature whereas the orbital fluctuations do not. The orbital fluctuations are less temperature dependent because of a rather large $U^\prime$ that suppresses them. In contrast the relatively small $J$ has the main impact on the spin fluctuations and therefore they set in at lower temperatures. A prominent feature of the ground state crossover is also the inversion of the orbital polarization, that agrees with our estimated critical value of $\tilde{t}^c_\perp$ in \refsec{sec:atomic-limit}

In order to estimate the evolution of quasiparticles we use the description of renormalized quasiparticle bands\cite{Najera2018}. The quasiparticle residue
\begin{equation}
  Z^{-1} = 1 - \frac{\partial \Re \Sigma(\omega)}{\partial \omega}\Big|_{\omega=0}
\end{equation}
renormalizes the non-interacting bandwidth $W=4t_b$ to
\begin{equation}
  \label{eq:zw}
  W_{\tilde{\epsilon}} = Z W
\end{equation}
and thereby the imaginary part of the self-energy $\Sigma(i\omega_n)$ on Matsubara axis encodes the coherence of the quasiparticles. Additionally, the real part of the self-energy shifts the energies of the quasiparticles
\begin{equation}
  \tilde{\epsilon} = Z\left(\tilde{t}_{loc} - \mu + \Re \Sigma(\omega = 0) \right)
\end{equation}
\begin{figure}
  \centering
  \includegraphics{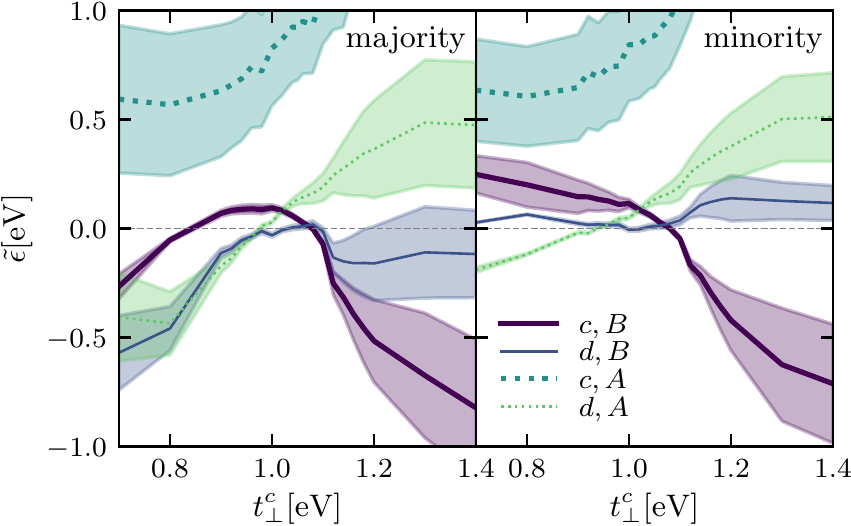}
  \caption{(Color online) Renormalized quasiparticle bands of the majority (left) and the minority (right) spin at $T = 0.02\eV$. Bonding (b) and antibonding (a) combinations of the atomic $c$ and $d$ orbitals. The renormalized bandwidths are represented by the colored regions.}
  \label{fig:qptcb50}
\end{figure}
One can see in \reffig{fig:qptcb50} that far from the critical region ($t^c_\perp \ll 1.05\eV$ or  $t^c_\perp \gg1.05\eV$) both $c$ and $d$ states are (mostly) shifted from the Fermi level (by strong bond-antibonding splitting and by correlation effects). In contrast there appears three  bands in the vicinity of the Fermi level close to critical $t^c_\perp$, which favours frustration effects.

In \reffig{fig:sztemptctd02} we use the notion of a quantum critical (QC) region for the low-temperature ($T\sim 0.02\eV$) paramagnetic phase. It is critical in a sense, that the scattering rates of all quasiparticles in proximity of the Fermi level diverge, i.e. the quasiparticle residue and renormalized bandwidth go to zero. The mechanism behind the formation of the paramagnetic insulator for $1\eV <t^c_\perp \leq 1.1\eV$  is the divergence of self-energies in several orbitals. This is distinct from interaction induced effective field splittings encoded in the real part of the self-energies and reminiscent of the Mott-insulator. The QC region is bounded from below. At low temperatures this criticality is avoided by the quasiparticles as they leave the Fermi level.

It is interesting that not only the realization of the different molecular spin-states (such as our DE and MO) has been observed experimentally in dimerized materials mentioned in Sec.~\ref{sec:methods} with general formula Ba$_3$MeTM$_2$O$_9$ depending of the choice of Me\cite{Ziat2017,Kimber2012,Senn2013a,Panda2015}. Moreover, some of these materials are characterized by a puzzling suppression of the long-range magnetic order and even a possible realization of the quantum spin-liquid phase due to frustrations~\cite{Terasaki2017,Rijssenbeek1998,Yamamoto2018}.

\section{Lattice effect}
The Bethe hopping parameter $t_b$ controls the embedding of the correlated dimer into the lattice. The limits of $t_b = 0$ and $t_b\rightarrow\infty$ correspond to isolated dimers and disconnected Bethe lattices, respectively. The situation of $t_b\ll t^c_\perp$, corresponds to not yet disconnected dimers, but ``uncorrelated'' ones with the charge concentrated on the bonds rather than sites. This state corresponds to the uncorrelated Peierls insulator. Apart from that $t_b$ controls the strength of quantum fluctuations of the bath, because it scales the hybridization for the corresponding Anderson impurity model that CDMFT maps to.

In this section we pick three values of $t_\perp^c=0.7\eV,1.05\eV,1.4\eV$ as representatives of the DE, HYB and MO states, respectively, at the temperature of $T=0.01\eV$ and vary the Bethe-hopping $t_b$ for each of them. The first part focuses on spin-polarized solutions and the second on paramagnetic ones.

\subsection{Dimer-antiferromagnetism}
\begin{figure}
  \centering
  \includegraphics{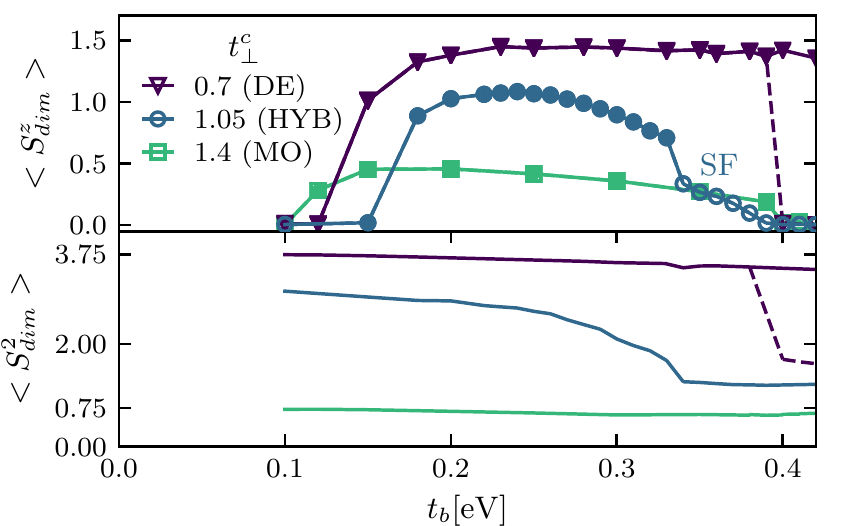}
  \caption{(Color online) Top: Dimer-magnetization $\expval{S^z_{dim}}$ as function of the Bethe-hopping $t_b$ for the dimer hoppings $t^c_\perp = 0.7\eV$ (DE), $1.05\eV$ (HYB), $1.4\eV$ (MO) at $T=0.01\eV$. At the crossover exists the spin freezing (SF) phenomenon at certain $t_b$. Filled and empty markers present insulating and metallic states, respectively. Metallicity is determined by analytical continuation using the maximum-entropy method. Bottom: Squared total-spin of the dimer $\expval{S_{dim}^2}$ as a function of the Bethe-hopping $t_b$ for the dimer hoppings $t^c_\perp = 0.7\eV,1.05\eV,1.4\eV$ at $T=0.01\eV$.}
  \label{fig:sz_s2_tb}
\end{figure}
The upper panel of \reffig{fig:sz_s2_tb} shows the dimer-magnetization $\expval{S^z_{dim}}$ (for simplicity we omit the $g$ factor) at $T=0.01\eV$  as a function of the Bethe hopping parameter $t_b$ in three regimes: the antiferromagnetically ordered DE ($t^c_\perp = 0.7\eV$) state, the crossover region ($t^c_\perp = 1.05\eV$) and the antiferromagnetically ordered MO ($t^c_\perp = 1.4\eV$) phase. One may see from this figure that there is no net magnetization in the limit of very small $t_b$ ($< 0.1\eV$), which corresponds to nearly isolated dimers as for $t_b=0.1\eV$ the single-particle gap of the $d$-orbital opens up.  In the region of intermediate $t_b$ both the DE and MO solutions have nearly maximal $\expval{S^z_{dim}}$, 3/2 and 1/2 respectively. It is interesting that the $t_b$ range of the nonzero magnetization is smallest for $t^c_\perp = 1.05\eV$ corresponding to the HYB state of the crossover region. Here, the fluctuations between the dimers are enhanced by the competing MO and DE states and suppress long-range magnetic order.

It is useful to compare upper and lower panels of  Fig.~\ref{fig:sz_s2_tb}, where square of the total spin, $\expval{S^2_{dim}}$, is plotted for the same values of $t^c_\perp$.  While $\expval{S^z_{dim}}$ measures ordered spin, $\expval{S^2_{dim}}$ simply tells us  what is the total spin of a configuration. The square of the total spin ($=S(S+1)$) for the DE and MO states in the atomic limit and at $T=0$ are 3.75 and 0.75. Comparing Figs.~\ref{fig:sz_s2_tb}(top) and \ref{fig:sz_s2_tb}(bottom) we, first make sure that two transitions for the MO solution at $t_b=0.1\eV$ and 0.4 are due to transition to the paramagnetic state, the total spin per the dimer is still well defined even for $t_b<0.1\eV$ and $t_b > 0.4\eV$. $\expval{S_{dim}^2}$ for both the MO and DE solutions  depend on $t_b$ only weakly. Thus, the formation of spin order is not due to local moment formation, it must be due to suppression of their fluctuations.

Second, we see from Fig.~\ref{fig:sz_s2_tb} that an increase of $t_b$ suppresses the DE state and increases the MO contribution in the crossover region (i.e. for $t^c_\perp = 1.05\eV$). Using corresponding values of $\expval {S^2_{dim}}$ for these two states one may estimate their contributions to the wavefunction for arbitrary $t_b$. If for $t_b=0.1\eV$ there is roughly 50/50 ratio between weights of the DE and MO states, then for $t_b=0.35$ we have $\sim$90\% of the MO and only 10\% of the DE state. This can be rationalized by treating it with a correction to the total energy of both states due to hopping within Bethe lattice, i.e. $t_b$, using the perturbation theory. 

We assume that the intra-dimer hopping $t^c_\perp$, Hubbard $U$, and Hund's exchange $J$ are leading parameters. Then the second order correction to the total energy due to $t_b$ would be $\sim -t_b^2/\delta \varepsilon$, where $\delta \varepsilon$ is the energy difference between excited and ground states. Clearly, the excited energy for the MO state will be much smaller than for the DE configuration, since the transfer of the $d$ electrons between two antiferromagnetically coupled dimers in the MO state does not cost Hund's exchange energy (there are two electrons with opposite spin projections on the bonding $c$ orbitals in the MO state and transferring $d$ electrons between dimers we keep the number of electrons (per site) with the same spin). Neglecting spin-flip and pair-hopping terms for simplicity we get  $\delta \varepsilon_{MO} \sim U/2$. The transfer of the $c$ electrons in the MO configuration is rather unfavorable, since it is possible only to antibonding orbitals. In contrast one may transfer the $c$ electrons in the DE state, this gives $\delta \varepsilon_{DE} \sim U/2 + J$, while an electron hopping via $d$ orbitals results in $\delta \varepsilon_{DE} \sim U + J/2$ - both much larger than energy of the excited state in the MO configuration. This explains the gradual increase of the MO weight and the decrease of $\expval{S^2_{dim}}$ in the crossover region with increasing $t_b$.

Third, there is a rather non-trivial evolution of both $\expval{S^z_{dim}}$ and $\expval{S^2_{dim}}$ with $t_b$ for $t^c_\perp = 0.7\eV$ (i.e. nominally for the DE solution). In particular for large $t_b$ ($\geq 0.4\eV$) we observe coexistence of two regimes: conventional insulating DE solution with long-range magnetic ordering and $\expval{S^z_{dim}}=3/2$, and the second one - metallic and paramagnetic with suppressed $\expval{S_{dim}^2} \approx 2$. The value of $\expval{S_{dim}^2}$ for the second solution is close to what one may expect for the spin-triplet.

\begin{figure}
  \centering
  \includegraphics{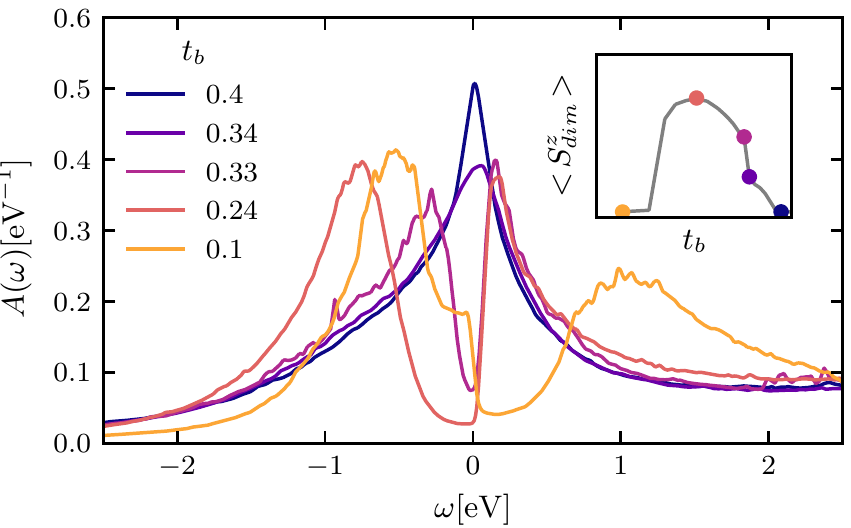}
  \caption{(Color online) The dimer density of states for different Bethe hoppings $t_b$ at $T=0.01$ and $t^c_\perp = 1.05$ obtained via the stochastic optimization method\cite{Krivenko2018,Goulko2017}. Inset: The corresponding dimer magnetizations $\expval{S^z_{dim}}$.}
  \label{fig:dos_freeze}
\end{figure}
\reffig{fig:dos_freeze} shows the local density of states in the crossover region with increase of $t_b$. One can see that for $t_b = 0.4\eV$ our system is in a metallic state, characterized by a large quasiparticle peak. Reducing $t_b$ we come to the broken spin symmetry situation, where the peak becomes less coherent (the width at half maximum height decreases), and then eventually observe formation of a pseudo-gap for $t_b=0.33\eV$, which corresponds to a sudden increase of the magnetization. The maximum of $\expval{S^z_{dim}}$ is exceeded at $t_b \approx 0.24\eV$, where the pseudo-gap transforms to a real gap. A further decrease of $t_b$ results in a transition to the paramagnetic state, which is accompanied by a modification of the spectral function. In particular for $t_b=0.24\eV$ there is a sharp edge for electron excitations, while for $t_b=0.1\eV$ we have a sharp edge for hole excitations.

\begin{figure}
  \centering
  \includegraphics{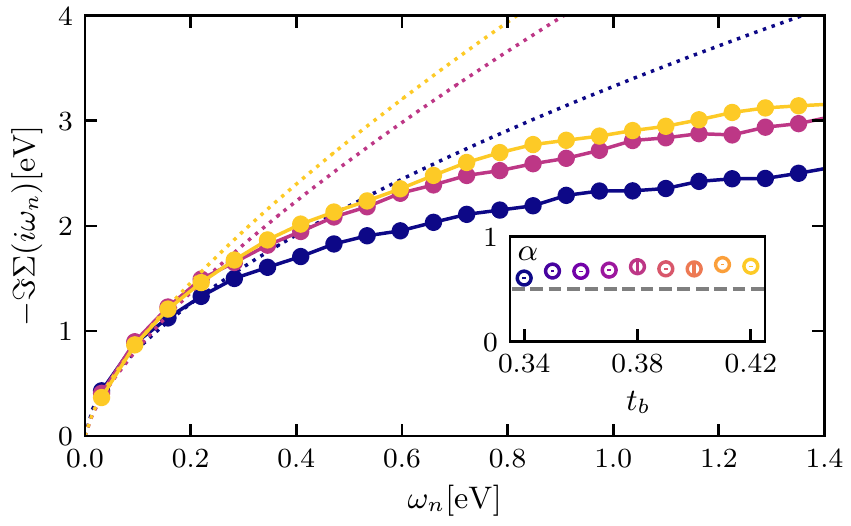}
  \caption{(Color online) Imaginary part of the self-energy (solid) on Matsubara frequencies together with a power-law low-frequency fit(dotted) for different $t_b$, $t^c_\perp=1.05\eV$ and $T=0.01\eV$. The fitted power is shown in the inset.}
  \label{fig:spinfreeze}
\end{figure}
In \reffig{fig:spinfreeze} we focus on the incoherent metal with local moments of $0.33\eV<t_b<0.4\eV$ and identify the underlying mechanism of spin-freezing which has been found in a previous single site DMFT multiorbital study at integer filling\cite{Werner2008,Stadler2015,Stadler2018} and is a property of Hund's metals. It is a non-Fermi liquid described by the constant spin-spin correlation function at long times and a strong enhancement of the local susceptibility\cite{Haule2009}. It has been pointed out that the ground state degeneracy seems to be an important component of spin-freezing. We can confirm that as our model shows the feature only in proximity of the ground state crossover. The self-energy of that phase is non-Fermi-liquid-like, but still the system is metallic in the freezing process. Since electrons scatter at the frozen moments, the self-energy shows power-law behavior $\Sigma(i\omega_n\rightarrow 0) = (i\omega_n)^\alpha$ with $\alpha < 1$ and can be fit with a quantum critical crossover function
\begin{equation}
  \label{eq:powerlaw}
  -\Im\Sigma(\omega_n)/t = C+A(\omega_n/t)^\alpha.
\end{equation}
A minimal exponent of $\alpha = 0.5$ was found at the critical point in the original study.\cite{Werner2008} At the magnetization jump, i.e. $t_b = 0.34\eV$, we also find a drop in $\alpha$ leading to a value $\alpha\approx 0.5$.

\subsection{Spectral properties}
Albeit a paramagnetic solution may be only metastable, one can enforce it to enhance scattering processes and thereby also amplify the electronic correlations. Thus, the paramagnetic solution is a tool to investigate ordering mechanisms and quasiparticles, whose diverging scattering rates eventually lead to a symmetry-broken solution.

\begin{figure}
  \centering
  \includegraphics{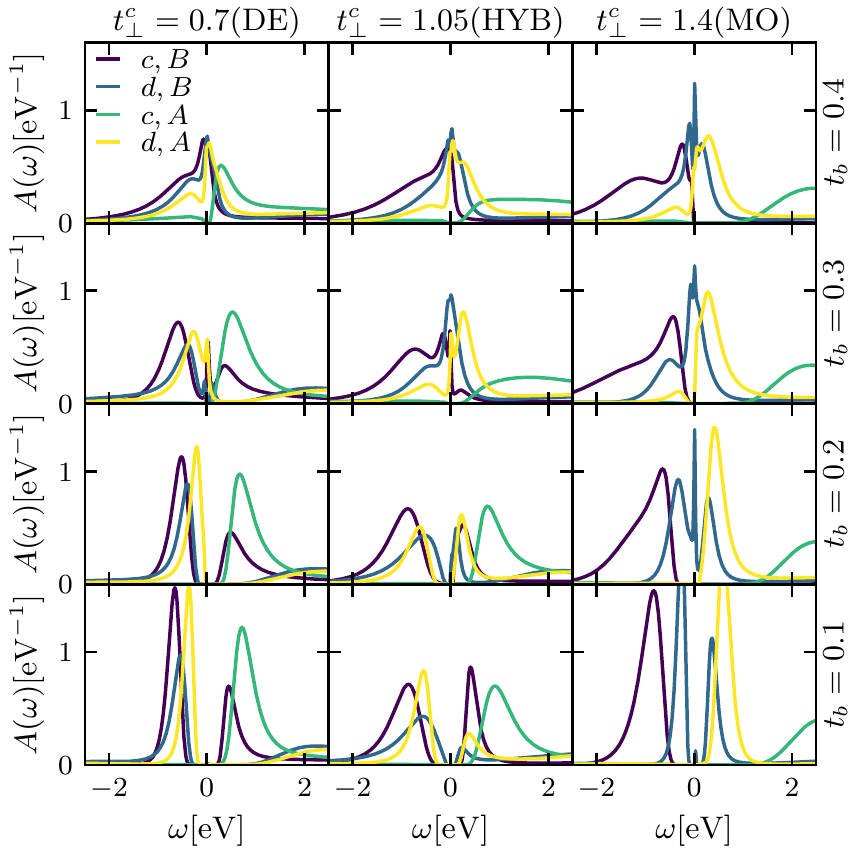}
  \caption{(Color online) Partial DOSes in BA representation as a function of intradimer hopping, $t^c_\perp$, and Bethe hopping, $t_b$, parameters (rows and columns, respectively). The fixed parameters are: $U=4.5\eV$, $J=0.7\eV$, $t^d_\perp= 0.2\eV$, $T = 0.01\eV$ and paramagnetism is enforced. Obtained by the Maximum Entropy method.}
  \label{fig:pm_dos_vs_tctb}
\end{figure}
\begin{figure}
  \centering
  \includegraphics{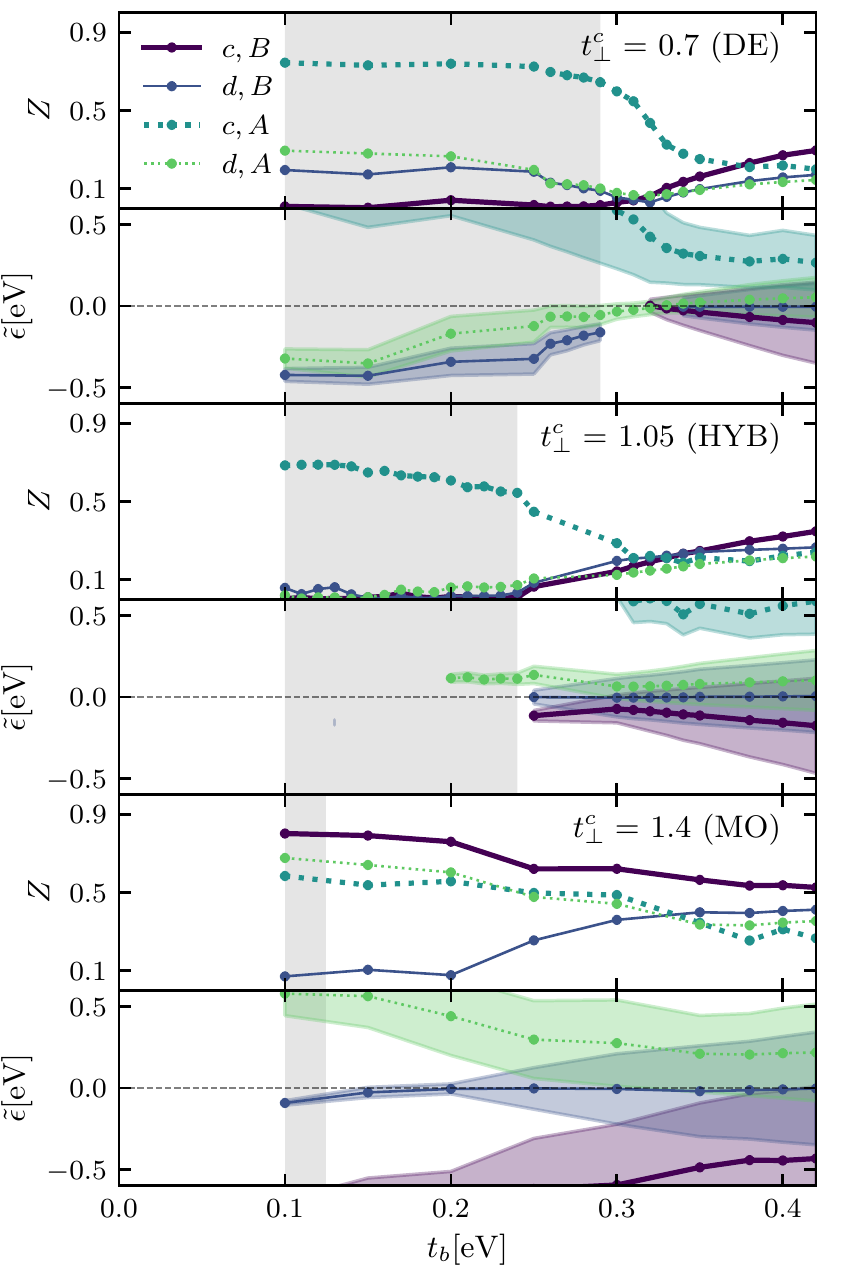}
  \caption{(Color online) Quasiparticle residues $Z$ and renormalized quasiparticle bands $(\tilde{\epsilon}, W_{\tilde{\epsilon}})$ as function of the Bethe-hopping $t_b$. The renormalized quasiparticle bands vanish if $Z\approx 0$. The shaded area depicts insulating phases determined by analytical continuation (maximum entropy method). Paramagnetism is enforced, $T = 0.01$.}
  \label{fig:zeqp}
\end{figure}
\reffig{fig:pm_dos_vs_tctb} presents partial DOSes in the BA representation for various values of the intradimer hopping of the $c$ electrons, $t^c_\perp$, and the Bethe hopping, $t_b$, that controls the bandwidth of non-interacting states. The most comprehensible is the MO state with $t^c_\perp=1.4 \eV$ and $t_b=0.1\eV$ (the lower-right part of the \reffig{fig:pm_dos_vs_tctb}). At these values of parameters, the bonding and antibonding $c$ orbitals are almost completely occupied ($n_{(c,B)}=1.78$) and empty ($n_{(c,A)}=0.2$), respectively, and can be integrated out. Therefore, one deals with a single electron in the double band model with crystal field splitting defined by $2t^d_\perp=0.4\eV$\cite{Poteryaev2008}. Such a large value of the crystal field splitting in comparison to the bandwidth, $W=4t_b=0.4\eV$, results in a further lifting of the degeneracy, and finally, one has a conventional Mott-Hubbard single band insulator, which occurs for the ($d,B$) orbital. By increasing $t_b$ (from bottom to top, right column of \reffig{fig:pm_dos_vs_tctb}) this insulating state evaluates to a single band metal ($t_b=0.2\eV$) and further to a three band metal at $t_b=0.4 \eV$. The latter happens due to such factor as bandwidth increasing of ($c,B$) and ($d,A$) states and its touching of the Fermi level (see lower panel of \reffig{fig:zeqp}). One should note that the ($c,A$) state remains empty at all values of the Bethe hopping. This picture of the insulator to metal transition is confirmed by the renormalized quasiparticle bands, $(\tilde{\epsilon}, W_{\tilde{\epsilon}})$, and the quasiparticle residue $Z$, shown in \reffig{fig:zeqp} (lower panel). The energetic order of the bands is determined by $t_{loc}$, i.e. bonding orbitals are lower than antibonding and the $c$ orbital is lower than the $d$ orbital. At small values of $t_b$, all renormalized bands except ($d,B$) are placed far from the Fermi level. The corresponding quasiparticle residue, $Z_{(d,B)}$, is close to zero. At $t_b>$0.2 eV  $Z_{(d,B)}$ is increased and system becomes a correlated  metal.

The spectral function of the DE state (lower-left part of the  Fig.~\ref{fig:pm_dos_vs_tctb}) is also consistent with the atomic picture. The $(d,B)$ and $(d,A)$ states are occupied with one electron per spin-orbital, $n_{(d,B)}=n_{(d,A)}=1$, that is equivalent to a single electron occupation of site centered orbitals. The remaining electron is on the $(c,B)$ state (antibonding part is completely empty). The Coulomb interaction leads to a gap opening for these states in different ways. Although the QP bands for all these orbitals are away from the Fermi level (see upper panel of the Fig.~\ref{fig:zeqp}), the quasiparticle residues behave differently for $(c,B)$ and $(d,B)$,$(d,A)$ states. $Z_{(c,B)}$ goes to zero at small values of $t_b$ while the larger $t_b$ they have finite values. It results in the orbital selective Mott transition at increased values of $t_b=0.3\eV$. A further increase of $t_b$ closes the gap in the $(c,B)$ spectral function. One should note that the overall quasiparticle residues of the DE solution are smaller then its MO counterparts indicating stronger electronic correlations in this regime.

The hybrid state, $t^c_\perp = 1.05\eV$, has an even stronger quasiparticle renormalization for all orbitals than the DE state. The $(c,B)$, $(d,B)$ and $(d,A)$ quasiparticle residues go to zero approximately at $t_b=0.25\eV$. This is related to the quantum critical region, that we have discussed in the context of \reffig{fig:sztemptctd02}. It results in the metal to insulator transition and gap opening in the corresponding spectral functions, see \reffig{fig:pm_dos_vs_tctb}.

It is interesting to note that a critical value of the Bethe hopping, $t^*_b$ decreases with the increase of the intradimer hopping parameter $t^c_{\perp}$. In the MO case there is only one active electron, which leads to an increased value of the critical Coulomb interaction for the multiband model~\cite{Georges2013}, that also corresponds to a decreased value of $t_b$. With the decrease of the intradimer hopping, $t^c_{\perp}$, all of the electrons have to be regarded for a description of the model. Therefore, the effective number of electrons is raised, that results in the decreased value of the Coulomb interaction parameter, or increased value of the critical Bethe lattice hopping, $t_b$.

\section{Conclusions}
We have studied the molecule formation of $3/8$-filled multiorbital dimers in solids as a consequence of the alignment of the octrahdra that form the ligands of the transition metal oxides and its impact on the dimer-antiferromagnetic order. We have chosen the parameters motivated by density functional calculations on the materials Ba$_3$YRu$_2$O$_9$ and Y$_5$Mo$_2$O$_{12}$ as representatives for the DE and MO states, respectively. They differ by the overlap of the $c$ electrons $t_\perp^c$. The estimated condition separating these two states was found to be $\tilde{t}^c_\perp = (3U-13J)/4$.

Close to that parameter the lattice of such dimers is in a strongly correlated state, that shows a paramagnetic region that is extended with respect to temperature and thereby the suppression of AFM order. It exhibits electrons with strongly renormalized masses in both orbitals ($c$ and $d$) and separates orbital decoupling from spin decoupling. Both decouplings originate from the dimer-groundstate crossover of $\tilde{t}^c_\perp$. Furthermore, the long-range spin order is more sensitive to temperature fluctuations than the orbital order and that renders the MO state more stable against temperature fluctuations.

Correlation effects could be induced by the change of the electron's itineracy within the Bethe-planes ($t_b$) as it promotes quantum fluctuations from the lattice on the dimers. We explain the larger stability of the AFM order in the MO configuration with the exchange energy $t_b^2/(U/2)$ of the $d$ electrons as opposed to the exchange energy of the DE state, i.e. $t_b^2/(U/2+J)$ for the $c$ electrons. The competition of the DE and MO states causes the formation of a new hybrid state, that exhibits qualitatively new features, e.g. an incoherent  metallic spin-polarized state with a non-Fermi liquid self-energy corresponding to the spin-freezing phenomenon.

We used the cluster DMFT to study the correlation enhanced enforced paramagnetic calculations that unveiled the orbital-selectiveness of the DE state -- typical for Hund's physics. The MO state shows correlation features, but the metal to insulator transition is rather of Peierls-type. Finally the hybrid state has a metal insulator transition involving the renormalization of all $d$ and the bonding $c$ states around the same value of the Bethe lattice hopping emphasizing the large impact of competing interactions on the electronic correlations.

\begin{acknowledgments}
M.H. and A.L. gratefully acknowledge financial support by the Deutsche Forschungsgemeinschaft (DFG) in the framework of the SFB 925. The work of S.V.S. and A.I.P. was supported by the Russian president science council via MD 916.2017.2 program, by the Russian ministry of science and high education via theme ``spin'' AAAA-A18-118020290104-2, ``electron'' AAAA-A18-118020190098-5, and contract 02.A03.21.0006, and by the Ural Branch of Russian academy of science (18-10-2-37). The computations were performed with resources provided by the North-German Supercomputing Alliance (HLRN).
\end{acknowledgments}

\bibliography{movsdebib,sergey}

\end{document}